\documentclass{emulateapj}

\usepackage{latexsym}  % for XeMulate style
\usepackage{epsfig}
\usepackage{graphics}
\usepackage{natbib}
\usepackage{color}
\usepackage{amsmath}
\usepackage{hyperref} 
\newcommand{\id}{{{\rm d}}}
\newcommand{\Xe}{{x_{\rm e}}}

\newcommand{\Yp}{{Y_{\rm p}}}
\newcommand{\Neff}{{N_{\rm eff}}}
\newcommand{\Ne}{n_{\rm e}}

\slugcomment{submitted to ApJ}
\shorttitle{recombination history from damping tail}

\shortauthors{Farhang et al.}

\begin{document}

\title{Constraints on perturbations to the recombination history from \\measurements
of the CMB damping tail}

\author{M.~Farhang\altaffilmark{1,2}, J.R.~Bond\altaffilmark{1},  J.~Chluba\altaffilmark{1,3}, E.R.~Switzer\altaffilmark{1}}
\altaffiltext{1}{Canadian Institute for Theoretical Astrophysics, 60 St George , Toronto, ON, M5S 3H8.}
\altaffiltext{2}{ Department of Astronomy and Astrophysics, University of Toronto, 50 St George , Toronto, ON, M5S 3H4.}
\altaffiltext{3}{Johns Hopkins University, Bloomberg Center 435 ,3400 N. Charles Street, Baltimore, MD, 21218.}

\begin{abstract}
The primordial CMB at small angular scales is sensitive to the ionization and expansion history of the universe around the time of 
recombination. This dependence has been exploited to constrain the helium abundance and the effective number of relativistic species. 
Here we focus on allowed ionization fraction trajectories, $\Xe (z)$,  by constraining low-order principal components of perturbations to the standard recombination scenario ($\Xe$-eigenmodes)  in  the circa 2011 SPT, ACT and WMAP7 data. Although the trajectories  are statistically consistent with the standard recombination, we find that there is a tension similar to that found by  varying the helium fraction. 
As this paper was in press, final SPT and ACT datasets were released and we applied
our framework to them: we  find the tension continues,  with slightly higher
significance, in the new 2012 SPT data, but find no tension with the standard model
of recombination in the new 2012 ACT data.
We find that  the prior probabilities on the eigenamplitudes  
 are substantially influenced by the requirement that $\Xe$ trajectories conserve electron number.  We propose requiring a sufficient entropy decrease between posterior and prior marginalized distributions be used as an $\Xe$-mode selection criterion. 
We find that in the case of the  2011 SPT/ACT+WMAP7 data only two modes are constrainable, but upcoming ACTPol, Planck and SPTPol data will be able to test more modes and more precisely  address the current tension.
\end{abstract}
%%*****************

\section{Introduction}
%--------------------
A primary goal of CMB experiments after the discovery of the acoustic features in the angular power spectrum was to measure the damping tail. This was achieved by the CBI \citep{cbi03,cbi07} and ACBAR  \citep{acbar03,acbar09} experiments, and spectacularly verified by the Atacama Cosmology Telescope \citep[ACT, e.g.,][]{act} and the South Pole Telescope \citep[SPT, e.g.,][]{spt}. There is even higher precision data on its way from ACT, SPT and Planck. 
Initially the goal was simply to confirm that the damping tail agreed with theoretical predictions based on cosmological parameters determined from the first set of peaks and troughs. Contemporary high-precision data opens the opportunity for new constraints of cosmological parameters that specifically influence the damping tail and have little influence on larger scales. The parameters wagged the tail, and now the tail can wag the parameters.  Examples of these parameters are  the primordial helium abundance $\Yp$ and the effective number of relativistic species, $\Neff$. Both experiments have hinted at a deficit of power in the damping tail through the  $\Yp$  and $\Neff$ probes \citep{spt,act,hou11}. Other physical possibilities have also been investigated to explain this damping tail tension; these include dark radiation \citep{smi12,egg12}, the annihilation or decay of dark matter particles \citep{Giesen2012}, cosmic strings \citep{liz12}  or a high-frequency cosmic gravitational wave background \citep{sen12}. 

CMB anisotropies are suppressed on small scales by shear viscosity and thermal diffusion when the photons and electrons are tightly coupled, and by higher order transport effects as the photons break out from their random-walk Thomson scatterings by electrons to approach free-streaming. The physics of Silk damping \citep{sil68} has been heavily explored, as reviewed in \cite{bond96,hu97,cbi03}, and is very accurately computed numerically in CMB transport codes \citep{CMBFAST,CAMB}. One can associate a characteristic damping wavenumber $k_{\rm D}(z)$ with the processes, determined by the steep variation in the mean free path $(\Ne (z) \sigma_{\rm T})^{-1}$  with redshift, where $\Ne (z)$ is the free-electron number density at redshift $z$ and $\sigma_{\rm T}$ is the Thomson scattering cross section. The overall effect on the CMB power spectrum $C_\ell$ can be approximately characterized by an exponentially damped envelope suppressing the baryon acoustic oscillations, encoded in an associated angular multipole damping scale $\ell_{\rm D} \sim k_{\rm D} (z_{\rm dec}) \chi (z_{\rm dec}) \sim 1350$, where $\chi(z_{\rm dec})$ is the comoving distance to the last scattering  redshift $z_{\rm dec}\sim 1088$. 
 
 Our focus here is how the damping tail is impacted by perturbations of the free-electron fraction, $\Xe(z)=\Ne/(n_{\rm p}+n_{\rm HI})$, where $n_{\rm p}$ and 
$n_{\rm HI}$ are the number density of ionized and neutral hydrogen atoms, during cosmological recombination in a model-independent way. This complements model-independent studies of $H(z)$ variations \citep{sam12}. Opening the constraint to the functional form of $\Xe(z)$ broadens the range of non-standard physics that can be tested, such as energy injection from decaying particles or dark matter annihilation \citep{che04, pad05}, which cause a delay of recombination \citep{Peebles2000}. 

The present data are only suggestive of a damping deficit, so discrimination between physical causes is not yet possible, but we develop a framework for considering general modifications and the impact of physical requirements on the posterior distributions. The recombination history is a central quantity in the interpretation of CMB data, and it has evolved significantly over time as more and more effects have been included \citep{Zeldovich68, Peebles68, sea99}, and, with extensive recent work,  it may now have nearly converged \citep{chl11, Yacine2010c}. It is important, however, to test directly whether tensions may exist which suggest that there are missing elements to the recombination story. For example, residual uncertainties in $\Xe(z)$ directly affect our ability to distinguish between different inflationary scenarios \citep{Hu1995, Lewis2006, rub10, sha11}. 
 
In this paper we follow the approach of \cite{far12} to generate eigenmodes for perturbations to the high redshift ionization history focussing on the epoch of recombination. The method is applied to the
seven-year WMAP data combined with the 2011 SPT, ACT data and their recent updates,  using a Fisher matrix eigenanalysis (Section~\ref{sec:method}) of finely binned redshift-localized $\Xe$-modes. In \cite{far12} the eigenanalysis was performed for simulated datasets consisting of CMB maps (or their spherical harmonic amplitudes $a_{\ell m}$). 
Here, the modes are defined from the measured CMB bandpower errors, which leads to a different structure for the Fisher matrix. 
Further extending Farhang et al. 2012, we jointly treat the foreground nuisance parameters of the ACT, SPT, and WMAP7 experimental data.
We use these data to constrain the amplitude of deviations from the standard recombination scenario in Section~\ref{sec:results}.  In Section~\ref{sec:priorspace}, we  investigate the impact of imposing electron number conservation (through $\Xe(z)$) on the initial and final probability distributions of the parameters and quantify its effect relative to the information delivered by the data. We conclude with a brief discussion in Section~\ref{sec:disc}. 
%

%----------------------------------------
\section{Eigenmodes for Perturbations to the High-z Ionization History}\label{sec:method} 
%----------------------------------------------
\subsection{Eigenmode construction}
%-----------------------------------------------
The functional form of deviations from the standard recombination history can be decomposed into a set of uncorrelated functions ranked by their significance in the data, here the increasing order of their forecasted errors. We construct these functions from the inverse of the Fisher information matrix ${\bf F}$ for the given combination of data used, 
\begin{equation}
{\bf F}_{ij}   =\sum_{b, b'}\frac{\partial C_{b}}{\partial q_i} {\rm Cov}_{bb'}^{-1}\frac{\partial C_{b'}}{\partial q_j}, \label{eq:fisher}
\end{equation}
where the $C_b$'s are the simulated bandpower measurements for a fiducial set of cosmic parameters and for a given experiment, and ${\rm Cov}_{bb'}$ is the bandpower covariance matrix for that experiment. The parameters $q_{i}$ include recombination perturbation parameters as well as the standard cosmological parameters and the various nuisance parameters for the experiments and for secondary effects. This form for 
${\bf F}_{ij}$ assumes uniform prior distributions for the parameters, Gaussian likelihoods for the bandpower data points and that information on the parameters mainly comes from the mean of the bandpowers, $C_b$,  rather than their covariance, ${\rm Cov}$. The latter can be verified for the high multipoles of interest by comparing the contributions to the Fisher matrix  from  the mean and the  covariance of data  \citep[see, e.g.,][]{teg97}.  Gaussianity  is a reasonable approximation again for the high multipoles which are of interest in this work (see, e.g., \cite{ver03}). Further, it is the  assumption adopted in the released likelihood codes for SPT and ACT. 

We use $\delta \ln(\Xe(z))$ as the perturbation parameter.    
$\Xe$ is the parameter of direct relevance to probe the atomic physics involved at recombination. It is also straightforward to limit its variations within physical ranges. 
However, it does not affect the CMB anisotropies as closely as, e.g., $\Ne$ and $\tau$ do. Using $\Ne$ would also decrease possible degeneracies with baryonic matter density.
Using the log in the expansion balances the low and high $z$-regimes, but the data define the region of dominant impact, namely around decoupling, rather solidly in the hydrogen recombination regime. 
These  perturbations can be represented interchangeably by any basis that represents their degrees of freedom and does not produce numerical errors    \citep[see][for a discussion on other parametrizations as well as various extended and localized basis functions]{far12}. Here we have used the cubic $M_4$ spline \citep[see e.g.][]{mon05} representation, and confirmed convergence against an increasing number of basis functions, for the explicit Fisher form, eq.~\ref{eq:fisher}, for these experiments.

As mentioned above,  the Fisher matrix considered here includes the $\Xe(z)$ amplitudes, the instrument-dependent nuisance parameters and the six primary cosmological parameters $(\Omega_{\rm b}h^2,\Omega_{\rm dm}h^2,\theta_{\rm s},\tau, n_{\rm s},\Delta^2_\mathcal{R})$, respectively describing the physical baryon density, physical dark matter density, the angular size of sound horizon at the last scattering surface, the reionization optical depth, the scalar spectral index,  and the curvature fluctuation amplitude. In taking the full Fisher inverse and focussing on sub-blocks of it, we are in effect marginalizing over the parameter directions not in the block in the approximation of a Gaussian posterior. 

To analyze multiple experiments with different sets of nuisance parameters, their individual Fisher matrices should be constructed and marginalized to give effective matrices containing only standard cosmological and $\Xe$ perturbation parameters which are common between all experiments. The effective Fisher matrices, now with the same dimension, are added to get the total Fisher matrix $\tilde{\bf F}$. The $\Xe$ perturbation eigenmodes are the eigenvectors of the perturbation block of  ${\tilde{\bf F}}^{-1}$ with their uncertainties estimated from the roots of the corresponding eigenvalues, assuming a Gaussian distribution for the mode amplitudes in the vicinity of their maximum likelihoods. Henceforth, the $\Xe$ eigenmodes are unambiguously referred to as the eigenmodes or simply the {\it modes}. 
Details on the Fisher eigenanalysis for multiple experiments with different nuisance parameters are discussed in Appendix~\ref{app:fisher}. To generate the bandpowers required in the Fisher matrix construction we modified the publicly available code {\sc Camb}\footnote{http://camb.info/}\citep{CAMB} to include the more general ionization histories required for this work.

%---------------------------------------------------------
\subsection{Datasets and their eigenmodes}\label{sec:eXeM}
%----------------------------------------------------------
Throughout this work, the SPT \citep{spt} and ACT \citep{act} measurements of the CMB temperature and the seven-year WMAP measurement of CMB temperature and polarization \citep{wmap7} are used to study perturbations around the ionization history of the universe during recombination.\footnote{The small overlap of the observed regions of SPT and ACT telescopes with WMAP has been neglected in this work.}  The SPT and ACT data consist of observation of $790$ and $296$ ${\rm deg}^2$ of the sky, at $150$ and $148$ GHz, during 2008- 2009 and 2008 seasons, respectively. For simplicity, we neglect non-CMB cosmological constraints. 

\begin{figure*} 
\begin{center}
\includegraphics[scale=0.5]{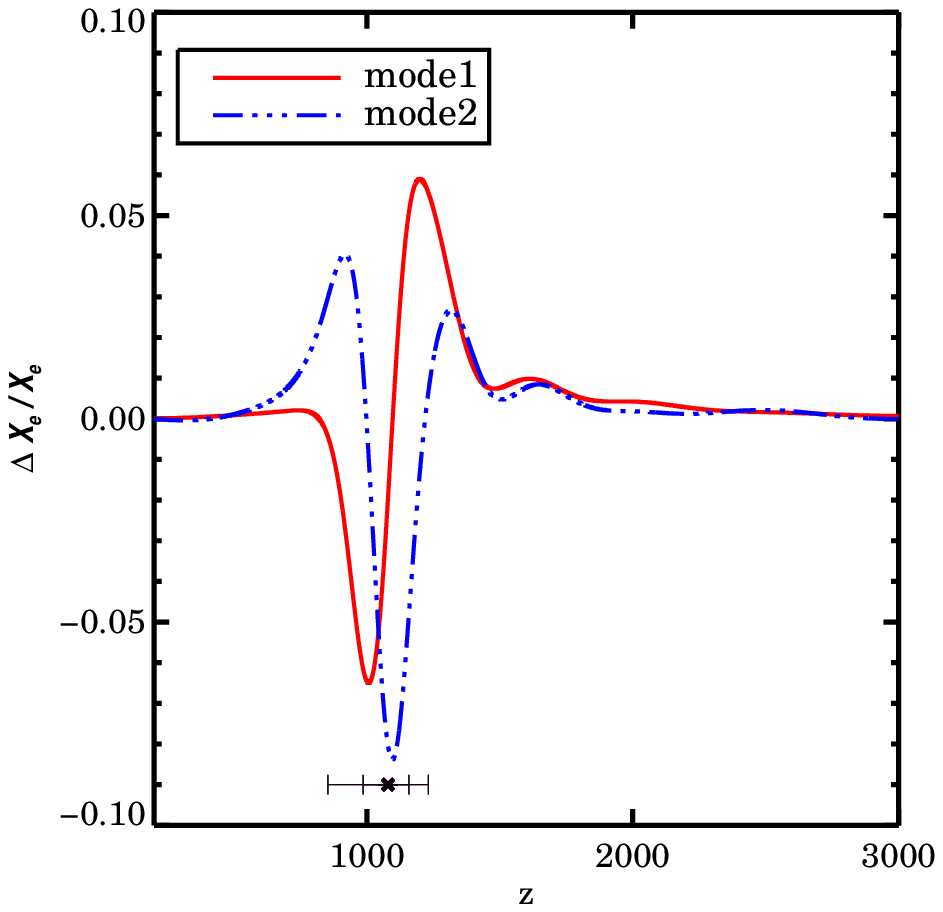}
\includegraphics[scale=0.5]{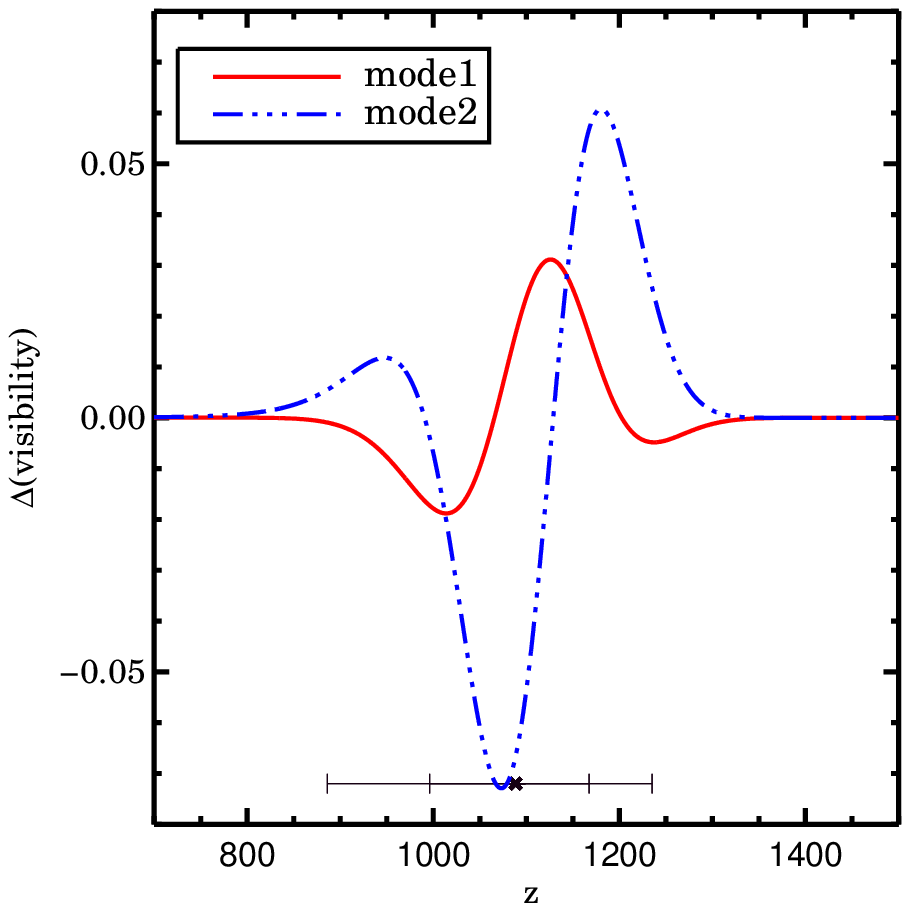}
\includegraphics[scale=0.5]{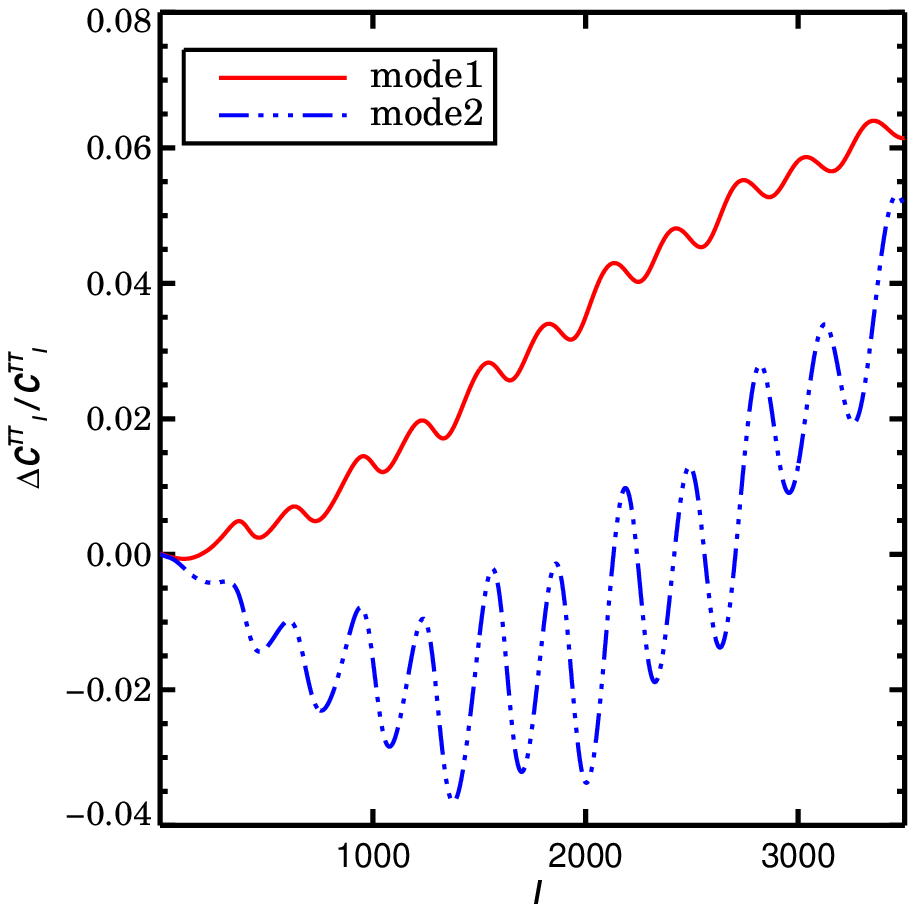}
\end{center}
\caption{The first two modes (constructed for SPT+WMAP7 data), normalized to have unit norm (left), the corresponding changes in the visibility functions (middle), and the resulting differential changes in the temperature power spectrum ${\rm C}_\ell$ (right). 
The middle and right plots corresponds to perturbations with SNR=1. 
The visibility function is defined as $g(z)=\id e^{-\tau}/\id\eta$, where $\eta$ is the conformal time and $\tau$ is the optical depth to the last scattering surface. The visibility functions have been normalized to the maximum of the fiducial model's visibility, which occurs at $z_{\rm dec}=1088$.  The width (at $68\%$ and $95\%$ levels) of the visibility function has been 
marked as error bars about $z_{\rm dec}$ in the figures. }
\label{Xpmnt-eXeM}
\end{figure*}

The left plot in Figure~\ref{Xpmnt-eXeM} shows the first two ionization eigenmodes generated for the combined SPT+WMAP7 dataset. The right and middle plots show the impact of these modes (with an amplitude equal to their $1\sigma$ error bars) on the CMB temperature power spectrum and on the Thomson differential visibility $g(z)=\Ne \sigma_{\rm T} c (1+z)^{-1} e^{-\tau}$, where the Thomson depth to redshift $z$ is $\tau (z) = \int \Ne \sigma_{\rm T} /H {\rm d}\ln (1+z)$.  The first two ACT+WMAP7 modes look very similar, so for definiteness, unless stated otherwise, {\it modes} refer to these and lower significance SPT+WMAP7 eigenmodes in this paper. Indeed, it turns out that the first two $\Xe$ perturbation modes for Planck+ACTPol data and for a high resolution cosmic variance limited experiment also have similar shapes \citep{far12}, with their order reversed in some cases.  We now show that the physical significance of the dominant first mode is, not surprisingly, intimately related to basic perturbative features in the differential visibility.

%-----------------------------
%----------------------------

\subsection{The Damping Physics of the Low Order Recombination Modes} \label{sec:recombdamp}

Since the rank-ordered eigenmodes are direct probes of the map from $\Xe$-trajectories to ${\rm C}_\ell$, the data-sensitive top-ranked modes should reflect the most basic ${\rm C}_\ell$-sensitive recombination effects, namely through the damping tail, which is intimately related to the sharply-peaked differential visibility.  We find the first two SPT+WMAP7 modes confirm this: they largely describe shifts in the  decoupling redshift (defined as the peak of $g(z)$,  $z_{{\rm dec}} \approx 1088$) and shifts in the decoupling width, $\sigma_{z,{\rm dec}}$,  the ``1-sigma" spread in $g$. We find that  a $+1\sigma$ amplitude for  the first mode changes the visibility by $\sim - 1.4\%$ in the width and by $\sim 0.4\%$ in the position of the peak;  for the second mode, the width increases by $\sim 6.2\%$ and the peak by $\sim 0.8\%$. 

The physical processes that define the structure of the damping tail have been well understood for a long time \citep[for a review see ][and references therein]{bond96,hu97}, and were discussed in relation to the experimental unveiling of the damping tail, first by CBI \citep{cbi03} and then by ACBAR \citep{acbar03}. Not surprisingly, the tail is controlled by the Compton scattering rate, $\Ne\sigma_{\rm T} c$, and the way it runs as the baryon density $n_{\rm b}$ drops, characterized by the local power law index, $p_{\rm e}=3{\rm d}\ln (\Ne /n_{\rm b} )/{\rm d}\ln n_{\rm b}$ $ = {\rm d}\ln \Xe /{\rm d}\ln (1+z)$ \citep{bond96}. The basic recombination quantities can be related to $p_{\rm e}$, which is zero at low and high $z$, has a maximum of about 12 and is about $9$ at $z_{\rm dec}$ for $\Lambda$CDM. The peak of the differential visibility $g(z)$ occurs when $\Ne\sigma_{\rm T}/H(z) = p_{\rm e}+2$, and the ``Gaussian'' width of decoupling in $ \ln (1+z)/\ln (1+z_{\rm dec}) $ is  $\sigma_{z,{\rm dec}} \sim (p_{\rm e}+2)^{-1}$. Thus for $\Lambda$CDM, the Compton time is about 1/20 of the horizon size at $z_{{\rm dec}}$, about 1/5 of the sound crossing time,  
and the relative width is about 0.09.   

Earlier than decoupling, the photons and baryons are so tightly coupled by Thomson scattering that they can be treated as a single fluid with sound speed $c_{\rm s}= (1+R)^{-1/2} c/\sqrt{3}$, lowered by the extra inertia of the baryons, $R \equiv {3 \bar{\rho}_{\rm b}
\over 4\bar{\rho}_\gamma}$, a photon+baryon kinematic shear viscosity $ (4/5) c_{\rm s}^2 (\Ne\sigma_{\rm T} c)^{-1}$ (in a full treatment of Thomson
scattering including angular anisotropy and polarization effects), zero bulk viscosity, and thermal
conductivity $\kappa_{\gamma} = n_{\rm b} s_\gamma (\Ne\sigma_{\rm T} c)^{-1}$, where $s_\gamma \sim 10^{9.8}$ is the photon entropy per baryon.
In this tightly coupled regime, a WKB treatment of the perturbed photon density shows the baryon acoustic oscillations are exponentially damped, $\propto \exp (-\int \Gamma /H {\rm d}\ln a )$, where the  sound wave damping rate relative to the Hubble expansion rate is \citep[see section C.3.1 in][]{bond96,kai83}
\begin{eqnarray}
&&\Gamma /H = {1\over 2} (kc_{\rm s} /Ha)^2 {H \over \Ne\sigma_{\rm T} c} {16\over 15}  \left[1+{\Gamma_{\rm diff}\over \Gamma_{\rm visc}} \right],   \nonumber \\
&& {\Gamma_{\rm diff}\over \Gamma_{\rm visc}}  = {15R^2\over 16 (1+R)}  \, . \label{eq:GoverH}
\end{eqnarray} 
Here $a=(1+z)^{-1}$. $kc_{\rm s} /Ha$ multiplies the comoving wavenumber $k$ of the acoustic oscillations by (approximately) 
the comoving distance sound travels in a Hubble time, $c_{\rm s} (Ha)^{-1}$. The  contribution of thermal diffusion  relative to that of shear viscosity is $\Gamma_{\rm diff}/\Gamma_{\rm visc} \approx 0.22$. 

To relate this to a WKB damping 
envelope acting on ${\rm C}_\ell$, we replace $k$ by  $\ell/\chi_{{\rm dec}}$, where $\chi_{{\rm dec}}$ is the comoving distance from us to decoupling, and integrate  up to $\ln a_{{\rm dec}}$. The damping scale obtained is
\begin{eqnarray}
&&\ell_{\rm D} \approx 1.7 (p_{\rm e}+2) (1+z_{{\rm dec}})^{1/2} (c/{c}_{{\rm s, dec}}) \label{eq:ellD}\\
&& \times [1+\Gamma_{\rm diff}/\Gamma_{\rm visc}]^{-1/2} (1+a_{\rm eq}/a_{{\rm dec}})^{1/2}  \nonumber \\
&& \times  ({c}_{\rm s,dec}/\bar{c}_{\rm s,dec})\sqrt{1+{1/2 \over p_{\rm e}+2}}\, . \nonumber
\end{eqnarray} 
The dominant first line gives the main dependences, $(1+z_{{\rm dec}})^{1/2}(p_{\rm e}+2)(1+R)^{1/2}$. With $p_{\rm e,dec} \approx 9$ and $z_{{\rm dec}} \approx 1088$, $c_{{\rm s,dec}}\approx 0.79 c/\sqrt{3}$ at decoupling, the first line gives $\ell_{\rm D} \sim 1360$. The terms in the second and third line are subdominant.  The first adds the thermal diffusion contribution to the viscous one,  giving a $\sim 10\%$ decrease; the second from the relativistic matter contribution to $H$ gives a $\sim 14\%$ rise. With these two, $\ell_{\rm D} \sim 1410$.  The fourth correction accounts for the decoupling sound speed being about $10\%$ lower than the speed averaged over the $z> z_{\rm dec}$ range; and the last $2.4\%$ correction occurs if we  use a sharp integration down to $z_{{\rm dec}}$, then stop. The third line terms change $\ell_{\rm D}$ to 1290, but are not there if we just replace $\int \Gamma/H {\rm d}\ln a $ by $\Gamma/H \sigma_{z,{\rm dec}}$. 

Of course a full transport treatment taking into account multipoles beyond the three which enter tight coupling (density, velocity and anisotropic stress) is required to get an accurate damping rate. The phenomenology adopted by \cite{hu97} estimated damping envelope functions multiplying ``undamped" acoustic ${\rm C}_\ell$'s from numerical ${\rm C}_\ell$-results, fitting them to a form $\exp[-(\ell /\ell_{\rm D})^{m_{\rm D}}]$, similar to the WKB approximation but with a floating slope to allow for a  slower falloff reflecting complexities beyond WKB physics (such as the less severe damping associated with  fuzziness of last scattering reflected in the $g(z)$ structure cf. the stronger viscous damping; the break-out into higher temperature multipoles in the Thomson-thick to Thomson-thin transition). For $\Lambda$CDM parameters, we get $\ell_{\rm D} \approx 1345$ and $m_{\rm D}\approx 1.26$, in better-than-expected  accord with the WKB estimate.  
 
 Apart from the residual memory of the acoustic oscillations, the rise in $\delta \ln {\rm C}_\ell$ of the first mode seen in Fig.~\ref{Xpmnt-eXeM}(c) conforms to the $(\ell /\ell_{\rm D})^{m_{\rm D}}$ form. The fluctuations in ${\rm C}_\ell$ are dominated by those in $\ell_{\rm D}$, with less sensitivity to $m_{\rm D}$. These are related to the fluctuations in the peak position and width (which is in turn related to $p_{\rm e,dec}$) by eq.~\ref{eq:ellD},
\begin{eqnarray*}
&&
\delta {\rm C}_\ell/{\rm C}_{{\rm f}\ell} \sim (\ell /\ell_{\rm fD})^{m_{\rm D}}  m_{\rm D} \delta \ell_{\rm D} /\ell_{\rm fD} \\
&& \delta \ell_{\rm D} /\ell_{\rm fD} \sim -{\delta \sigma_{z,{\rm dec}}\over \sigma_{{\rm f}z,{\rm dec}}} +{1\over 2} {\delta z_{{\rm dec}} \over  (1+z_{\rm  f,dec})} \, , 
\end{eqnarray*} 
with respect to the fiducial values with subscript $\rm f$. 
So we can interpret  the first, most significant, mode as primarily due to $\ell_{\rm D}$ variations. Similarly we can understand the sign change in $\Delta g$ for the first mode as being in response to $\delta z_{{\rm dec}}$. 

If we use the same approach for the influence of helium abundance fluctuations, the effect would be the smaller number of hydrogen nuclei near decoupling, suggesting a $\delta {\rm C}_\ell/{\rm C}_{{\rm f}\ell} \sim - (\ell /\ell_{\rm fD})^{m_{\rm D}}  m_{\rm D} \delta \Yp /(1-\Yp)$ form, in accord with what we see in Fig.~\ref{dCl2Cl_XNY}. 

%-------------------------------------------
\section{Constraints from circa 2011 ACT, SPT and WMAP7 data}\label{sec:results}
%--------------------------------------------
To search for perturbations in the standard recombination history ({\sc Recfast} \citep{sea99} with recent recombination corrections included \citep{chl11, Yacine2010c}, we use the amplitude of the modes introduced in Section~\ref{sec:eXeM} as a set of new parameters, and estimate them jointly with the six primary and three nuisance parameters. The nuisance parameters
follow \cite{spt} and \cite{act} and are the  amplitudes of the Poisson and clustered power from point sources, and a template for the total thermal and kinetic SZ power.   The shapes of the associated ${\rm C}_\ell$ templates do not look like our $\Xe$-modes, and the data can differentiate what is nuisance from what may be standard recombination deviation, albeit with correlations that are fully taken into account in the statistics. This statement holds even though we restricted ourselves to  single effective frequency analyses for SPT and ACT. The priors used for the nuisance parameters are taken from multi-band data particular to the flux cut for point source removal.  Unless stated otherwise, throughout the analysis, $\Yp$ and $\Neff$ are fixed to  $\Yp=0.2478$ \citep[from SPT data for the $\Lambda$CDM model and standard BBN,][]{spt} and $\Neff=3.046$ \citep[from the standard model of particle physics,][]{particlephysics12}.

 For parameter estimation we use the publicly available code, {\sc CosmoMC}\footnote{http://cosmologist.info/cosmomc/}, and modify it to include estimation of the mode amplitudes. We use the versions of {\sc CosmoMC} adapted for SPT\footnote{http://lambda.gsfc.nasa.gov/product/spt/spt\_spectra\_2011\_get.cfm} and ACT\footnote{http://lambda.gsfc.nasa.gov/product/act/act\_likelihood\_get.cfm} dataset likelihood functions. 
We checked that when the mode amplitudes are fixed to zero, the modified {\sc CosmoMC} recovers the reported SPT and ACT parameter measurements \citep{spt,act}. Lensing of the ${\rm C}_{\ell}$'s has been included throughout this work.

%-------------------------------------------------------------
\begin{deluxetable*}{c|ccc|ccc}
\tablecaption{The constraints on cosmological parameters with different sets of parameters used, as measured by SPT+WMAP7 and ACT+WMAP7. $\mu_1$ and $\mu_2$ refer to the amplitudes of the first and second modes.}
\tablehead{\colhead{} &\colhead{} & \colhead{SPT+WMAP7} & \colhead{ }  & 
  \colhead{ }  & \colhead{ACT+WMAP7}  & 
  \colhead{ }   \\
  \colhead{parameters} & \colhead{6s} & \colhead{$+$ mode 1}  & 
  \colhead{$+$  mode 2} & \colhead{6s} & \colhead{$+$ mode 1}  & 
  \colhead{$+$  mode 2}   
  }
  \startdata  
   $100 \Omega_{\rm b}h^2$    &  $2.221\pm 0.042$ & $2.253\pm 0.046$ & $2.249 \pm 0.047$  & $2.219 \pm 0.051 $ & $2.240\pm 0.050$      & $2.236 \pm 0.053$ \\[1mm] 
 $\Omega_{\rm c}h^2$      & $0.1110\pm 0.0048$ & $0.1123\pm 0.0049$ & $0.1118 \pm 0.0052$ & $0.1121 \pm 0.0052$ & $0.1155\pm 0.0056$   & $0.1121\pm 0.0061$  \\[1mm]  
 $100 \theta_{\rm s}$     & $ 1.041\pm 0.002$ & $1.041 \pm 0.002$ & $1.040 \pm 0.003 $ & $1.039 \pm 0.002$  & $ 1.039\pm 0.002$   & $1.035 \pm 0.004 $ \\[1mm] 
 $\tau$         & $0.086 \pm 0.015$ &  $0.089 \pm 0.015$ & $0.089 \pm 0.015 $   &$0.086\pm 0.015$ & $0.089 \pm 0.015$    & $0.0875\pm 0.015$  \\[1mm] 
 $n_{\rm s}$         &  $0.964 \pm 0.011$ & $0.977 \pm 0.013 $  & $ 0.975 \pm 0.016$  & $0.963 \pm 0.013 $ & $0.976 \pm 0.015$   &  $0.960\pm 0.019$ \\[1mm] 
 $10^9\Delta^2_\mathcal{R}$              & $2.43 \pm 0.10$ &  $2.40 \pm 0.10$ & $ 2.40 \pm 0.10$ & $2.45 \pm 0.11 $ & $2.43 \pm 0.11$     &  $2.45 \pm 0.11 $ \\[1mm] 
 $\mu_1\footnote{ The mode amplitudes and errors in this table (and throughout the paper) should be interpreted with respect to the normalized version of the modes as plotted in Figure~\ref{Xpmnt-eXeM}. So, e.g., perturbations with $\mu_1=1$ correspond to $\Xe$ changes in the form of mode 1 and with an amplitude exactly as plotted in Figure~\ref{Xpmnt-eXeM}.}$          &$(0)$ &   $-0.77 \pm 0.46$ & $-0.76 \pm 0.47$  &(0) &   $-1.27 \pm 0.74$   & $-1.67 \pm 0.86$ \\[1mm] 
 $\mu_2$         & $(0)$& $ (0) $ & $-0.39 \pm 1.09 $ & (0) &      $(0)$     & $-3.5 \pm 2.7$  \\[2mm] \hline

 $\sigma_8 $(derived)       & $0.807 \pm 0.024$& $ 0.825 \pm 0.027$ & $0.818 \pm 0.032$ & $0.814 \pm 0.028$& $ 0.841\pm 0.031 $ & $0.802 \pm 0.040$  \\[1mm]
 $\delta  z_{\rm dec}  / z_{\rm dec}  $ \footnote{relative change in the redshift of maximum visibility where $ z_{\rm dec}  =1088$ is the fiducial maximum visibility point.}& -- &  $-0.6\%$   &  $-0.7\%$  & -- &  $-1.0\%$   &  $-1.7\%$\\[1mm]
 $\delta \sigma_{z,{\rm dec}}/\sigma_{z,{\rm dec}}$ \footnote{relative change in the width of the visibility function.}& -- &  $1.5\%$  &  $-0.5\%$ & -- &  $2.6\%$  &  $-14.0 \%$ \\[1mm] 
 $(|\delta \Xe|/\Xe)_{\rm max}$\footnote{maximum relative change in the ionization fraction. The redshift corresponding to this maximum change is also included.} & -- &  $5\%~(z\sim 1196)$   &  $5\%~(z\sim 1039)$ & -- &  $8\%~(z\sim 1006)$   &  $31\%~(z\sim 1076)$\footnote{This large deviation, though looking curious, is not statistically significant. This point is understandable  given the relatively large estimated  values for  $\mu_1$ and $\mu_2$  and their uncertainties.} \\[1mm] 
 \hline
 $\Delta \chi^2 $    & -- & $ 2.5 $  &  $2.5$     & -- & $2.1$ & $2.5$
 \enddata
\label{bf-errors}
\end{deluxetable*}
%-------------------------------------------------------------
\begin{deluxetable*}{c|cc|cc}
\tablecaption{The constraints on the first two modes ($\mu_1$ and $\mu_2$), as measured by SPT12+WMAP7 and ACT12+WMAP7. }
\tablehead{\colhead{} & \colhead{SPT12+WMAP7} &  
  \colhead{ }  & \colhead{ACT12+WMAP7}  & 
  \colhead{ }   \\
  \colhead{parameters} &  \colhead{$+$ mode 1}  & 
  \colhead{$+$  mode 2} &  \colhead{$+$ mode 1}  & 
  \colhead{$+$  mode 2}   }
  \startdata  
 $\mu_1$  &   $-0.73 \pm 0.38$ & $-0.68 \pm 0.39$   &   $0.30 \pm 0.47$   & $-0.06 \pm 0.55$ \\ %[1mm] 
 $\mu_2$  & $ (0) $ & $-1.04 \pm 0.83 $ &       $(0)$     & $1.3 \pm 1.1$  %\\ 
  \enddata
\label{spt_act_recent}
\end{deluxetable*}
%-------------------------
Tables~\ref{bf-errors}  summarizes the results when one and two modes are used in the analysis, for SPT+WMAP7 and ACT+WMAP7 respectively, and compares them to the standard six-parameter model. Both experiments find a non-zero (but different) amplitude for the first mode, but they are only of   $1.7\sigma$ significance and so do not correspond to a detection; they have $\Delta \chi^2 \sim 2$. ACT+WMAP7 also has a non-zero mean for the second mode, but only at $1.3\sigma$. When the second mode is added to the ACT+WMAP7 analysis, the shift and uncertainty of the first mode  change slightly.  

While this paper was in press, improved measurements
of the damping tail were released by SPT \citep[][referred to as SPT12]{hou12} and ACT \citep[][referred to as ACT13]{dun13,sie13}. We present 
our new results using  these recent datasets for the SPT11+WMAP7 modes in Table 2.
We verified our SPT12 and ACT12 implementations in COSMOMC by checking we reproduce
the \cite{hou12} and \cite{sie13} determinations for the standard six cosmological
case and also for the 
cases including $\Yp$ or $\Neff$ variation. The table shows that the mode amplitudes for 
SPT12 are highly consistent with those of SPT11, with a slight error decrease, with
a mean for the 
first mode amplitude now $1.9\sigma$ away from zero.  As our discussion of the similarity
of the first mode $\delta C_\ell$ shape and the perturbed 
Helium abundance shape illustrates,  this is as expected from the new \cite{hou12} SPT measurement of  $\Yp = 0.300 \pm 0.025$ (and
$\Neff = 3.62\pm 0.48$). ACT12 reported values of these parameters that were consistent
with the unperturbed values of the basic-six parameter case, and as expected we see 
consistency with zero for the ACT12 mode amplitudes.

%-----------------------------------------
\section{The impact of electron number conservation on the posterior}
\label{sec:priorspace}
%--------------------------------------------
\begin{figure*} 
\begin{center}
\includegraphics[scale=0.7]{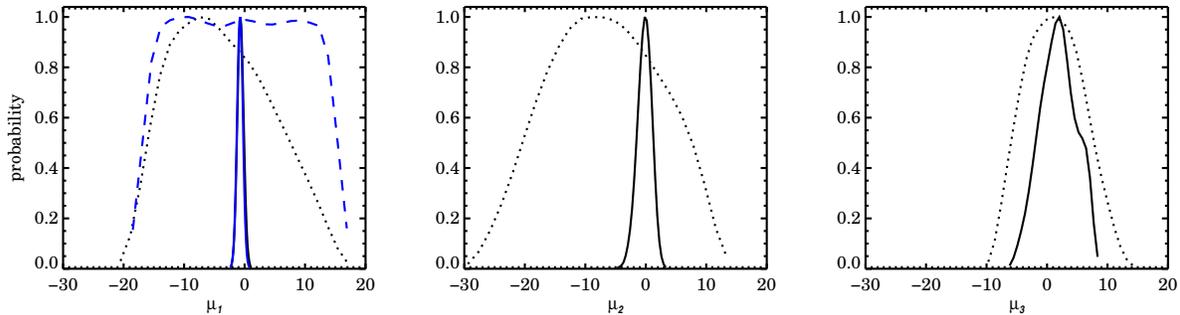}
\end{center}
\caption{Black lines: marginalized 1D-prior (dotted lines) and posterior probabilities (solid lines) with SPT+WMAP7 data for the first three modes, in an analysis where six standard parameters, three nuisance parameters and the first three $\Xe$ perturbation modes were used. Blue lines, left plot: the prior (dashed line) and posterior (solid line) distributions for $\mu_1$, similar to black lines, but with only one mode included in the analysis. Note that the solid black and blue curves coincide in the left plot. }
\label{pri-post-eXeM}
\end{figure*}
%-----------------------------------------------
Our perturbed ionization history is required to satisfy electron conservation through
%----
\begin{equation*}
0 \le \Xe(z) \le x_{\rm e}^{\rm max}
\end{equation*}
%--
where  $x_{\rm e}^{\rm max}=1+\frac{2\Yp}{3.97(1-\Yp)}$ is the maximum total electron fraction,  using $m_{\rm He}/m_{\rm H} \approx 3.97$, obtained when helium and hydrogen are fully ionized. When the mode amplitudes are poorly determined by data, the reconstructed $\Xe (z)$ could break through this bound, which of course we do not allow. Thus, although our starting prior may have been uniform with a wide possible spread in the amplitudes, the true prior distribution can only be determined with a full suite of Monte Carlo calculations restricting the allowed range. 
The Fisher analysis does  not catch this because the amplitudes are supposed to be tiny. They are not in our case for which the data do not have strong discriminatory power so allowed variations in the ionization history can be very broad. Intuitively, if the volume spanned by the prior space is comparable to or smaller than the volume of the likelihood space (at a given significance level), the posterior will be influenced by the physical constraint, and the Fisher matrix analysis will be a poor approximation to the full analysis.  Experiments with higher sensitivities will provide information about a larger number of modes before running into this condition, \citep[see simulations for Planck+ACTPol-like observations in][]{far12}.

%--------------------------------------------
 Figure~\ref{pri-post-eXeM} shows the marginalized $1$D-distributions of the amplitudes of the first three modes, $\mu_1$ to $\mu_3$, in an analysis with three modes included, and for two experimental setups: the posterior distributions of the SPT+WMAP7 case (solid black lines) and the prior-only simulations (dotted black lines) by ignoring the likelihood, i.e., assuming infinite errors in the data. 
The overplotted blue lines (in the left plot only) correspond to a case with only one mode being included in the analysis.
Note that the prior distributions of the first and second modes are skewed toward negative values for the case with three modes in the analysis (black lines). However, the prior distribution for the first-mode-only case (the quite symmetric blue dashed curve) shows that the $\mu_1$ measurement is not prior-driven. 

The very narrow posterior distributions of $\mu_1$ and $\mu_2$ relative to their priors illustrate the power of the ACT/SPT data in constraining these parameters, although they are not found to be significantly different from zero. For $\mu_3$, on the other hand, the comparable widths of the prior and posterior distributions imply that the dataset under consideration hardly contains more information  about this parameter than the limits set by electron conservation. The measured errors on the fourth and higher modes differ significantly from their Fisher forecasts, not even keeping their ranking. The insensitivity of the data to higher modes explains why we have limited our study to the first two modes. 

%-----------------------------------------
\begin{figure} 
\begin{center}
\includegraphics[scale=0.7]{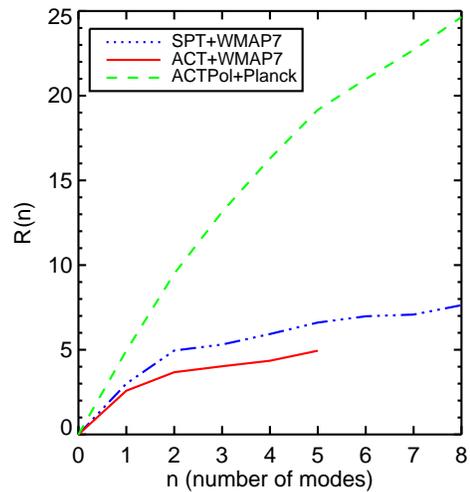}
\end{center}
\caption{Increase in the information content of the measured modes delivered by data relative to the volume of parameter space allowed by electron conservation, for different number of modes included in the analysis and various experimental cases. The modes in each case are constructed for the corresponding dataset.}
\label{Rn-spt-act-apol}
\end{figure}
%--------------------------------------------------
We can quantify the impact of the prior by measuring the Shannon entropy decrease in the measurement of $n$-parameters, ${\bf q}=\{q_i\}$, associated with the transition from the prior distribution $p_i$ to the 
posterior distribution $p_{\rm f}$ when data are added, 
%--------
\begin{equation}
R(n) \equiv S_{\rm i}(n)-S_{\rm f}(n) \equiv \langle {\ln p}_{\rm f} \rangle_{\rm f}  - \langle {\ln p}_{\rm i} \rangle_{\rm i}  \, . \label{eq:Rnfull}
\end{equation}
%-----------
For us the relevant $q_i$'s are the amplitudes of the first $n$ modes. 

Although the full calculation can be made, we have found that a Gaussian approximation works reasonably well, and does not have the numerical challenges associated with an accurate full calculation. The posterior $p_{\rm f}$ is closer to Gaussian than is the prior $p_{\rm i}$. With an $n$D-Gaussian distribution with zero mean and covariance matrices ${\bf C}$,  $S=1/2\ln {\rm det}({\bf C})+  n\ln (2\pi) /2 + 1/2 \langle {\bf q}^{\rm T} {\bf C} ^{-1}  {\bf q}\rangle $. The last term is $1/2$ of the mean $\chi^2$ associated with the measurement, hence is $n/2$ since  $\langle {\bf q}^{\rm T} {\bf C} ^{-1}  {\bf q}\rangle=n$. The entropy difference is then the ratio of mean log-volumes of the parameter space in question, namely 
%-------------
\begin{equation}
R(n)=\frac{1}{2}\ln ({\rm det}({\bf C}_{{\rm i}})/ {\rm det}({\bf C}_{{\rm f}})) , \label{eq:Rn}
\end{equation}
%-------------
where ${\bf C}_{{\rm i, f}}$ are the prior and posterior parameter covariance matrices.
We checked the Gaussian approximation  by comparing  eq.~\ref{eq:Rn} with estimates of the integral form eq.~\ref{eq:Rnfull}  of the entropies. The integral was calculated from the nearest neighbor entropy estimate \citep{sin03}. This non-parametric entropy estimation method is  based on the distribution of the nearest neighbor distance of the samples, here the MCMC chain outputs, and is  used for parameter spaces with more than one dimension.  We found that  the results from Gaussian approximation agree well with those from full integration, and are less noisy as the dimensionality of the parameter space increases. Using the determinant ratio to measure the level of improvement with improved data is  familiar as a figure-of-merit  \citep[see, e.g.,][]{mor10}. Although we have found for our application for deciding which modes to include that eq.~\ref{eq:Rn} is adequate, eq.~\ref{eq:Rnfull} is the better expression for a more accurate figure-of-merit \citep{far11}.
 
As $n$ increases, the data add less information about the parameters relative to the prior.  Thus, the difference between successive $R$'s gradually decreases. By adding new parameters, the volumes of the  posterior and prior spaces change by a similar prior-dominated factor. In the limit of very large $n$,  $R(n) \rightarrow {\rm constant}$. This is shown in Figure~\ref{Rn-spt-act-apol} where we compare  $R(n)$ for different  datasets and various numbers of modes included, $n$. For this plot, the modes of each curve are the eigenmodes constructed for the corresponding experiment.
 For ACT+WMAP7 and SPT+WMAP7, the difference between one and two parameters  is  greater than the difference between other subsequent modes. This shows that these datasets are much more informative about the second mode than the higher modes, which are entering the prior-dominated regime. This difference between the first two modes and the higher ones  is also evident from Figure~\ref{pri-post-eXeM}. With higher precision datasets, we expect the transition from likelihood to prior-dominance to happen at a higher mode number.  
   This prior-likelihood dominance transition hints to a natural criterion for mode-hierarchy truncation. However, one should note that  choosing a  quantitative mode selection criterion can be rather subjective and not necessarily applicable to all datasets.

The mode-selection criterion introduced here is much stronger than the Occam's razor argument developed in  \citet{far12}, where the truncation of the mode hierarchy was based on the change in the information content as more ordered modes were added.  Here, we have used the posterior information that the modes have in excess of the prior, a change of perspective motivated by the weak constraints from current datasets. The analysis enters a prior-dominated regime beyond just one or two modes and there is no need to consider the more sophisticated \cite{far12} criteria.
 
%------------------------------------------------
\section{Discussion}\label{sec:disc}
%--------------------------------------------------------
\begin{figure} 
\begin{center}
\includegraphics[scale=0.7]{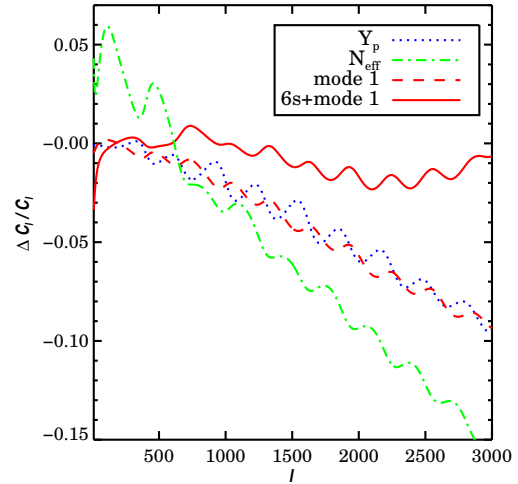}
\end{center}
\caption{ 
The solid red line corresponds to the difference between the best-fit ${\rm C}_\ell$'s for the standard case and the case with the first mode included, measured by SPT+WMAP7.  That is, the two cases have different background cosmology (as measured by data) as well as different values for $\mu_1$ ($\mu_1=0$ and $\mu_1=-0.77$). The dashed and dotted curves show the response to changes in $\Yp$, $\Neff$ and the first mode. 
In these cases, the six standard parameters are fixed for all models, while   
$\Yp=0.296$, $\Neff=3.898$ and $\mu_1=-0.77$ have been chosen for their corresponding curves.}\label{dCl2Cl_XNY}
\end{figure}
%--------------------------------------------

\begin{figure} 
\begin{center}
\includegraphics[scale=0.4]{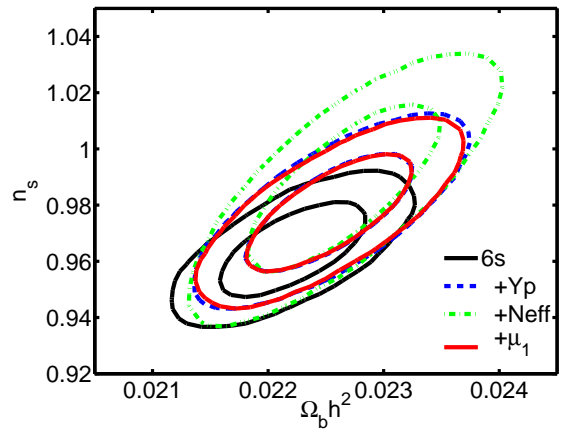}
\end{center}
\caption{ The marginalized $68\%$ and $95\%$ $n_{\rm s}$-$\Omega_{\rm b}h^2$ contours for  various sets of parameters being included in the analysis. The 6s contours represent the standard model with six parameters. Other cases each have one extra parameter, being $\Yp$, $\Neff$ and $\mu_1$. Note that these extended models favor a slightly higher value of $n_s$ compared to the standard case.}
\label{std_2D}
\end{figure}

In this work we studied how allowing for some freedom in the recombination history gives a better fit to the damping tail of the CMB power spectrum as measured by SPT and ACT, compared to the primary six-parameter model. The red solid line in Figure~\ref{dCl2Cl_XNY} shows the relatively small shift between the best-fit SPT+WMAP7 ${\rm C}_\ell$'s, one with the basic six parameters fixed at their best-fit values with the $\Xe$-perturbations on, and the other with the basic six parameters fixed at their unperturbed values. That is because the non-zero $\mu_1$ is accompanied by compensation in the values of the standard parameters, most significantly shifting $n_{\rm s}$ and $\Omega_{\rm b}h^2$ (see Figure~\ref{std_2D}) to give a small net $\Delta {\rm C}_\ell$. (Although the modes are marginalized over standard parameters, they are generally correlated with them.) When the basic six parameters are set to their $\Xe$-unperturbed values, and only $\mu_1$ varies, the red dashed line is obtained. Apart from the oscillation difference, the shape and value look rather like that for $\Yp$ variation, for the reasons discussed in Section~\ref{sec:recombdamp}. 

%--------------------------------------------------------
%--------------------------------------------

%--------------------------------------------------------
\begin{figure} 
\begin{center}
\includegraphics[scale=0.6]{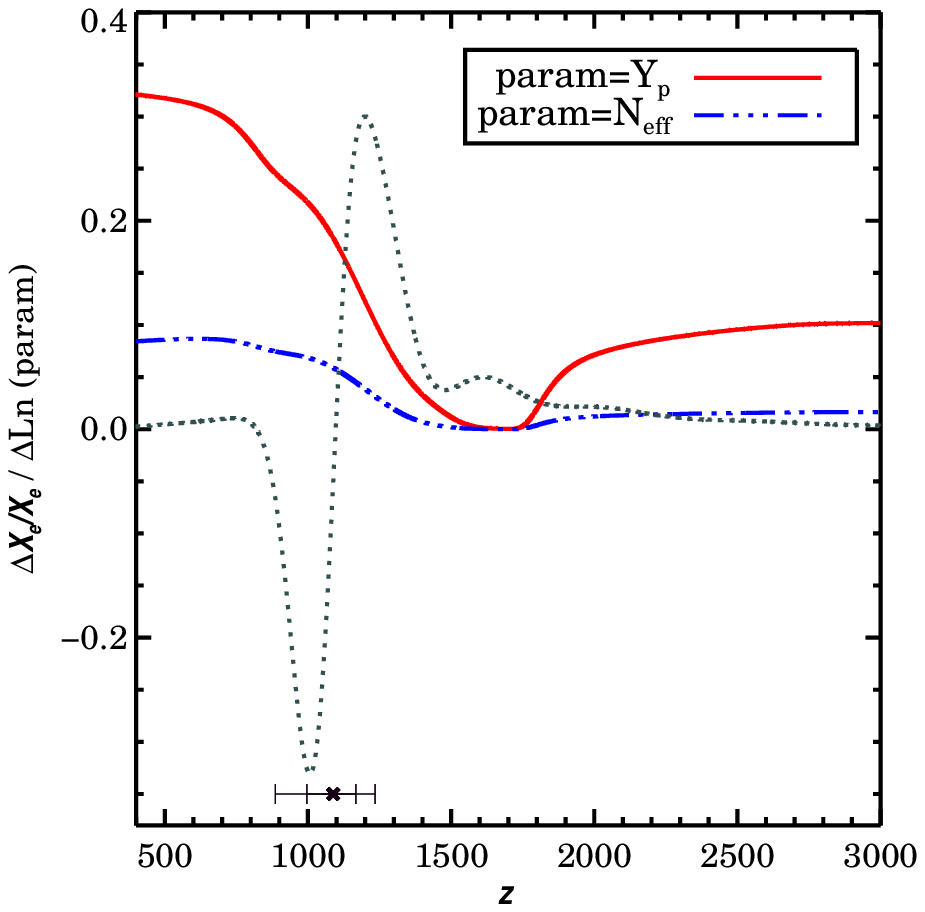}
\includegraphics[scale=0.6]{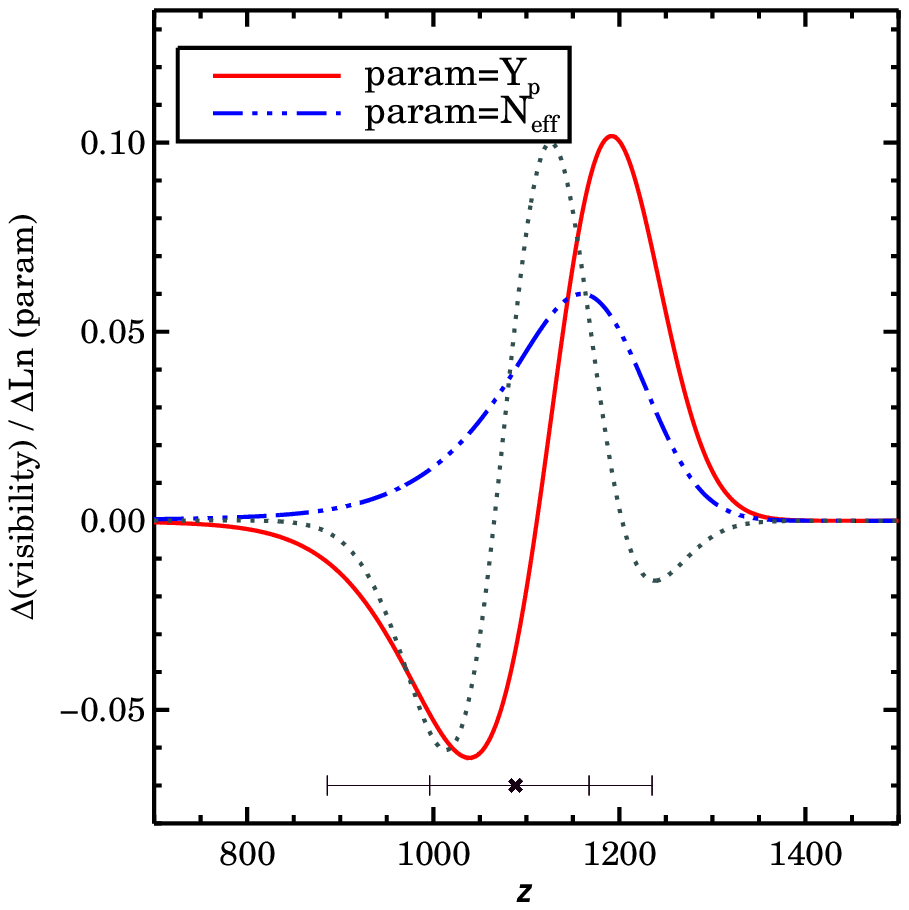}
\end{center}
\caption{The relative $\Xe$ and  differential visibility changes due to relative infinitesimal changes in $\Yp$ and $\Neff$ (with other parameters fixed). The first mode has been added in the background (the gray dotted lines) to aid visual comparison. It has been normalized to be comparable to $\Yp$ changes. }
\label{xe_vis_cl}
\end{figure}
%--------------------------------------------

The goodness of fit with $\mu_1$ added to the analysis is comparable to cases where the recombination history is assumed to be perfectly known \citep[see][for the most recent calculations]{chl11, Yacine2010c} and instead either $\Yp$ or $\Neff$ are allowed to vary \citep{spt,act}. 
(The corresponding  best-fit ${\rm C}_\ell$ difference between these two cases with the standard six parameter case is not shown here but is similar to the red solid line for the $\mu_1$ case.) As mentioned, including $\Yp$ has a similar  impact on the standard parameter measurements to that of $\mu_1$, as illustrated in Figure~\ref{std_2D}, where $n_{\rm s}$ and $\Omega_{\rm b}h^2$ have the most significant shifts. $\Neff$, on the other hand, behaves differently, with a different ${\rm C}_\ell$-shape,  and it also has a large impact on $\Omega_{\rm dm}h^2$ and $H_0$ \citep[see][]{spt,act}.

It is also noteworthy that the measurements of $\sigma_8$ (the amplitude of linear matter fluctuations at $z=0$ on scales of $8h^{-1} \rm{Mpc}$) for the model with $\Neff$ for ACT data \citep[$\sigma_8=0.906\pm 0.059$,][]{act} and for SPT data \citep[$\sigma_8=0.859\pm 0.043$,][]{spt} are currently slightly disfavored by $\sigma_8$ inferred from clusters \citep[$\sigma_8(\Omega_{\rm m}/0.25)^{0.47}=0.813 \pm 0.013 \pm 0.024$; here the second set of errors is systematic and due to the uncertainty in cluster masses,][]{vik09}. Including $\mu_1$ and especially $\mu_2$ in the analysis, with fixed $\Neff=3.046$, brings $\sigma_8$ towards lower values (see Table~\ref{bf-errors}), consistent with external datasets.

The first principal component $\Xe$-mode, $\Yp$, and $\Neff$ are strongly correlated due to their similar effects on the damping tail of the power spectrum as illustrated in Figure~\ref{dCl2Cl_XNY}. The correlation coefficients of $\mu_1$ with the other two are ${\rm corr}(\mu_1,\Yp)=0.95$ and ${\rm corr}(\mu_1,\Neff)=0.72$, based on results from {\sc CosmoMC} for SPT+WMAP7 bandpower data. Due to this partial degeneracy, varying more than one of these parameters simultaneously would significantly increase their uncertainties. The  improvement in errors forecasted for near-future data and the addition of polarization information will modestly reduce the degeneracy of $\mu_1$ with $\Yp$ and $\Neff$. For example,  for a Planck+ACTPol -like scenario the correlation coefficients for an $\ell$-by-$\ell$ analysis from {\sc CosmoMC}  are forecasted to  be ${\rm corr}(\mu_1,\Yp)=0.77$ and ${\rm corr }(\mu_1,\Neff)=0.60$. (An $\ell$-by-$\ell$ analysis is more sensitive to acoustic oscillation phase information than data with wide bandpowers.)

 The perturbative $\Xe(z)$-eigenmodes are, by definition,  those that are most constrained by the CMB. The perturbations in $\Xe (z)$ induced by varying $\Yp$ and $\Neff$, are shown in Fig.~\ref{xe_vis_cl}. They look very different than our most significant mode. What is interesting is the response in differential visibility. The first mode and the $\Yp$-induced perturbations look somewhat similar, although the differences are important. In ${\rm C}_\ell$ the dominant damping tail  behaviour of $\Delta {\rm C}_\ell$ is even closer, although the details of the peak-trough oscillations about it differ.  The full $\Yp$-induced perturbation involves a coherent sum over many eigenmodes, but the data are  mostly trying to constrain its component in the first mode, as Figure~\ref{dCl2Cl_XNY} shows. The low order  mode amplitudes are therefore a way to efficiently transfer information from the CMB into constraints on ionization perturbations.
Alternative high-$z$ ionization history models due to specific physical effects (and priors they may impose) could then be differentiated in e.g. the $\mu_1-\mu_2$ 
plane.

There are several alternative effects that can cause modifications to the CMB power spectrum in the damping tail.
These include possible modifications to the physics of recombination, dark radiation \citep[e.g.,][and references therein]{arc11}, changes in the fine structure constant $\alpha$ \citep{Kaplinghat1999, bat01, Scoccola2008}, high-frequency cosmic gravitational wave background \citep{smi06}, dark matter annihilation and particle decay \citep{che04, pad05, zha06, zha07, hue09, gal09, hue11, gal11, Giesen2012}. Deviations from standard recombination  may be differentiated or corroborated by non-CMB measurements. Apart from invoking additional physical processes to explain damping tail measurements, it is important to note that the tail is sensitive to experimental issues, such as  the instrumental beam,  point sources in the maps, and detector time constants. 

%The damping tail tension suggested by the ACT and SPT data should be detected or disproved with much higher significance by future Planck (to be released in 2013) and ACTPol/SPTPol data \citep{actpol,sptpol}. 
As shown in table~\ref{spt_act_recent}, we have now found the damping tail tension in the 
SPT11 and ACT11 data of table~\ref{spt_act_recent} persists  in the full $2500 ~{\rm deg}^2$  SPT data reported in 
 \cite{spt12a} and \cite{hou12}, but not in the three-season ACT12 data. 
The Planck data, with
its much larger sky coverage, should be able to address
whether tension exists or not. Our forecasts for Planck 2.5 year data show that 
 the current SPT11 best-fit amplitude of the
first mode,   $\mu_1 = -0.77$ (which is only $1.7\sigma$ now), could
be detected at more than $10\sigma$.

If such a deviation were found, it would be well beyond the levels of  the standard recombination corrections which have been discussed extensively over the past several years \citep[see][and additional references]{Dubrovich2005, Chluba2006, Kholu2006, Switzer2007I, Hirata2008, Grin2009, Yacine2010} and would have to indicate a new and possibly non-standard process at work. 
These physical mechanisms, with their  high level of nuance and theoretical concordance, result in a modification to recombination which is $3.5$ times smaller than the $1\sigma$ errors found here \citep{act}. 
In the far future, measurements of the cosmological recombination radiation from hydrogen and helium \citep[see][for an overview]{Sunyaev2009} may provide another way to investigate this question and break some of the expected degeneracies. 

If a detection of deviation is made with high significance, further explorations could be done with the same sort of analysis as that given here, but with modes weighted towards different redshift regimes.  Indeed, although we have focussed here on just the recombination epoch, viewing recombination and reionization as a connected $\Xe$-trajectory  has appeal, since CMB data (though at large rather than small angular scales) inform the latter. In these extended studies one needs to explore other possible variables to linearly expand in rather than $\delta \ln(\Xe(z))$, as discussed in \cite{far12}. The merit of   $\Xe(z)$ expansions is that one can weight them to concentrate on specific  recombination regions, e.g., at higher $z$ where helium recombines, or at lower $z$ as $\Xe$-freeze-out is approached. Ultimately showing in a model-independent way that the allowed $\Xe$-trajectories do not compromise our determination of cosmological parameters would  further increase our confidence in conclusions drawn from CMB datasets.

This work was accomplished by support from NSERC and the Canadian Institute for Advanced Research. 
%-----------------------------------------
 \appendix                                   
%----------------------------------------
\section{Fisher Analysis} \label{app:fisher}
The goal of this work is to search for deviations from the standard ionization scenario at high redshifts, around the epoch of recombination. For this purpose, we search  for the perturbation patterns in $\Xe$ best constrained by data \citep[see][for more details]{far12}. The most constrained perturbation parameters are the eigenvectors of the $\Xe$-perturbation block of the inverse of Fisher information matrix. 
The Fisher matrix for each dataset under consideration is 
%------
\begin{equation*}
{\bf F}_{ij}=\sum_{b,b'}\frac{\partial C^T_{b}}{\partial q_i} {\rm Cov}_{bb'}^{-1}\frac{\partial C_{b'}}{\partial q_j}
\end{equation*}
%-------
where the $q$'s represent any of the standard, the $\Xe$-perturbation, the secondary or experimental nuisance parameters.  
The bandpowers $C_b$ are
%-
\begin{equation*}
C_b=\sum W_{b\ell} \tilde{C}_{\ell} ~,~~ \tilde{C}_{\ell} = \frac{\ell(\ell+1)}{2 \pi} {\rm C}_\ell
\end{equation*}
 where the window functions, $W_{b\ell}$'s, are specific to the experiment and ${\rm Cov}_{bb'}=\langle\delta C_b \delta C_{b'}\rangle $ is their covariance matrix. The derivatives are calculated at the fiducial $q_i$'s.
 If multiple experiments are used, the total Fisher matrix is the sum of the individual Fisher matrices constructed for each experiment --- if they are statistically independent --- and marginalized over their nuisance, experiment-dependent parameters.
To make this marginalized Fisher matrix, we divide ${\bf F}$ into blocks of nuisance parameters (represented by $n$) and the cosmologically interesting parameters, represented by $y$. We then have
 \begin{align*} 
\bf{F}&=
\left( {\begin{array}{cc}
{\bf F}_{yy} & {\bf F}_{yn}\\
{\bf F}_{ny} & {\bf F}_{nn}\\
\end{array}} \right),  ~~~~~
{\bf F}|_{m}={\bf F}_{yy}-{\bf F}_{yn}{\bf F}^{-1}_{nn}{\bf F}_{ny}
\end{align*}
%-------
where ${\bf F}|_{m}$ is the Fisher matrix (including only cosmologically interesting parameters) marginalized over nuisance parameters.
The individual ${\bf F}|_{\rm m}$'s should be added  to get the total Fisher matrix.
The $\Xe$ perturbation eigenmodes are the eigenvectors of the perturbation block of the total Fisher matrix after it has been marginalized over the standard parameters, similar to the above marginalization. 

Another approach is to calculate Fisher matrices that include all nuisance parameters. Then, for those experiments that provide no constraint on a set of nuisance parameters, set those matrix entries to zero, sum the matrices over experiments and marginalize over nuisance and standard cosmic parameters.

%*****************
\bibliography{xe_data}

\begin{thebibliography}{64}
\expandafter\ifx\csname natexlab\endcsname\relax\def\natexlab#1{#1}\fi

\bibitem[{{Ali-Ha{\"i}moud} \& {Hirata}(2010)}]{Yacine2010}
{Ali-Ha{\"i}moud}, Y., \& {Hirata}, C.~M. 2010, \prd, 82, 063521

\bibitem[{{Ali-Ha{\"i}moud} \& {Hirata}(2011)}]{Yacine2010c}
---. 2011, \prd, 83, 043513

\bibitem[{{Archidiacono} {et~al.}(2011){Archidiacono}, {Calabrese}, \&
  {Melchiorri}}]{arc11}
{Archidiacono}, M., {Calabrese}, E., \& {Melchiorri}, A. 2011, \prd, 84, 123008

\bibitem[{{Battye} {et~al.}(2001){Battye}, {Crittenden}, \& {Weller}}]{bat01}
{Battye}, R.~A., {Crittenden}, R., \& {Weller}, J. 2001, \prd, 63, 043505

\bibitem[{Beringer {et~al.}(2012)Beringer, Arguin, Barnett, Copic, Dahl, Groom,
  Lin, Lys, Murayama, Wohl, Yao, Zyla, Amsler, Antonelli, Asner, Baer, Band,
  Basaglia, Bauer, Beatty, Belousov, Bergren, Bernardi, Bertl, Bethke, Bichsel,
  Biebel, Blucher, Blusk, Brooijmans, Buchmueller, Cahn, Carena, Ceccucci,
  Chakraborty, Chen, Chivukula, Cowan, D'Ambrosio, Damour, de~Florian,
  de~Gouv\^ea, DeGrand, de~Jong, Dissertori, Dobrescu, Doser, Drees, Edwards,
  Eidelman, Erler, Ezhela, Fetscher, Fields, Foster, Gaisser, Garren, Gerber,
  Gerbier, Gherghetta, Golwala, Goodman, Grab, Gritsan, Grivaz, Gr\"unewald,
  Gurtu, Gutsche, Haber, Hagiwara, Hagmann, Hanhart, Hashimoto, Hayes, Heffner,
  Heltsley, Hern\'andez-Rey, Hikasa, H\"ocker, Holder, Holtkamp, Huston,
  Jackson, Johnson, Junk, Karlen, Kirkby, Klein, Klempt, Kowalewski, Krauss,
  Kreps, Krusche, Kuyanov, Kwon, Lahav, Laiho, Langacker, Liddle, Ligeti, Liss,
  Littenberg, Lugovsky, Lugovsky, Mannel, Manohar, Marciano, Martin, Masoni,
  Matthews, Milstead, Miquel, M\"onig, Moortgat, Nakamura, Narain, Nason,
  Navas, Neubert, Nevski, Nir, Olive, Pape, Parsons, Patrignani, Peacock,
  Petcov, Piepke, Pomarol, Punzi, Quadt, Raby, Raffelt, Ratcliff, Richardson,
  Roesler, Rolli, Romaniouk, Rosenberg, Rosner, Sachrajda, Sakai, Salam,
  Sarkar, Sauli, Schneider, Scholberg, Scott, Seligman, Shaevitz, Sharpe,
  Silari, Sj\"ostrand, Skands, Smith, Smoot, Spanier, Spieler, Stahl, Stanev,
  Stone, Sumiyoshi, Syphers, Takahashi, Tanabashi, Terning, Titov, Tkachenko,
  T\"ornqvist, Tovey, Valencia, van Bibber, Venanzoni, Vincter, Vogel, Vogt,
  Walkowiak, Walter, Ward, Watari, Weiglein, Weinberg, Wiencke, Wolfenstein,
  Womersley, Woody, Workman, Yamamoto, Zeller, Zenin, Zhang, Zhu, Harper,
  Lugovsky, \& Schaffner}]{particlephysics12}
Beringer, J., {et~al.} 2012, Phys. Rev. D, 86, 010001

\bibitem[{{Bond}(1996)}]{bond96}
{Bond}, J.~R. 1996, in Cosmology and Large Scale Structure, ed. {R.~Schaeffer,
  J.~Silk, M.~Spiro, \& J.~Zinn-Justin}, 469--+

\bibitem[{{Chen} \& {Kamionkowski}(2004)}]{che04}
{Chen}, X., \& {Kamionkowski}, M. 2004, \prd, 70, 043502

\bibitem[{{Chluba} \& {Sunyaev}(2006)}]{Chluba2006}
{Chluba}, J., \& {Sunyaev}, R.~A. 2006, \aap, 446, 39

\bibitem[{{Chluba} \& {Thomas}(2011)}]{chl11}
{Chluba}, J., \& {Thomas}, R.~M. 2011, \mnras, 412, 748

\bibitem[{{Dubrovich} \& {Grachev}(2005)}]{Dubrovich2005}
{Dubrovich}, V.~K., \& {Grachev}, S.~I. 2005, Astronomy Letters, 31, 359

\bibitem[{{Dunkley} {et~al.}(2011){Dunkley}, {Hlozek}, {Sievers}, {Acquaviva},
  {Ade}, {Aguirre}, {Amiri}, {Appel}, {Barrientos}, {Battistelli}, {Bond},
  {Brown}, {Burger}, {Chervenak}, {Das}, {Devlin}, {Dicker}, {Bertrand
  Doriese}, {D{\"u}nner}, {Essinger-Hileman}, {Fisher}, {Fowler}, {Hajian},
  {Halpern}, {Hasselfield}, {Hern{\'a}ndez-Monteagudo}, {Hilton}, {Hilton},
  {Hincks}, {Huffenberger}, {Hughes}, {Hughes}, {Infante}, {Irwin}, {Juin},
  {Kaul}, {Klein}, {Kosowsky}, {Lau}, {Limon}, {Lin}, {Lupton}, {Marriage},
  {Marsden}, {Mauskopf}, {Menanteau}, {Moodley}, {Moseley}, {Netterfield},
  {Niemack}, {Nolta}, {Page}, {Parker}, {Partridge}, {Reid}, {Sehgal},
  {Sherwin}, {Spergel}, {Staggs}, {Swetz}, {Switzer}, {Thornton}, {Trac},
  {Tucker}, {Warne}, {Wollack}, \& {Zhao}}]{act}
{Dunkley}, J., {et~al.} 2011, \apj, 739, 52

\bibitem[{{Dunkley} {et~al.}(2013){Dunkley}, {Calabrese}, {Sievers}, {Addison},
  {Battaglia}, {Battistelli}, {Bond}, {Das}, {Devlin}, {Dunner}, {Fowler},
  {Gralla}, {Hajian}, {Halpern}, {Hasselfield}, {Hincks}, {Hlozek}, {Hughes},
  {Irwin}, {Kosowsky}, {Louis}, {Marriage}, {Marsden}, {Menanteau}, {Moodley},
  {Niemack}, {Nolta}, {Page}, {Partridge}, {Sehgal}, {Spergel}, {Staggs},
  {Switzer}, {Trac}, \& {Wollack}}]{dun13}
---. 2013, ArXiv e-prints

\bibitem[{{Eggers Bjaelde} {et~al.}(2012){Eggers Bjaelde}, {Das}, \&
  {Moss}}]{egg12}
{Eggers Bjaelde}, O., {Das}, S., \& {Moss}, A. 2012, ArXiv:1205.0553

\bibitem[{{Farhang} {et~al.}(2012){Farhang}, {Bond}, \& {Chluba}}]{far12}
{Farhang}, M., {Bond}, J.~R., \& {Chluba}, J. 2012, \apj, 752, 88

\bibitem[{{Farhang} {et~al.}(2011){Farhang}, {Bond}, {Dor{\'e}}, \& {Barth
  Netterfield}}]{far11}
{Farhang}, M., {Bond}, J.~R., {Dor{\'e}}, O., \& {Barth Netterfield}, C. 2011,
  ArXiv e-prints

\bibitem[{{Galli} {et~al.}(2009){Galli}, {Iocco}, {Bertone}, \&
  {Melchiorri}}]{gal09}
{Galli}, S., {Iocco}, F., {Bertone}, G., \& {Melchiorri}, A. 2009, \prd, 80,
  023505

\bibitem[{{Galli} {et~al.}(2011){Galli}, {Iocco}, {Bertone}, \&
  {Melchiorri}}]{gal11}
---. 2011, \prd, 84, 027302

\bibitem[{{Giesen} {et~al.}(2012){Giesen}, {Lesgourgues}, {Audren}, \&
  {Ali-Ha{\"i}moud}}]{Giesen2012}
{Giesen}, G., {Lesgourgues}, J., {Audren}, B., \& {Ali-Ha{\"i}moud}, Y. 2012,
  ArXiv:1209.0247

\bibitem[{{Goldstein} {et~al.}(2003){Goldstein}, {Ade}, {Bock}, {Bond},
  {Cantalupo}, {Contaldi}, {Daub}, {Holzapfel}, {Kuo}, {Lange}, {Lueker},
  {Newcomb}, {Peterson}, {Pogosyan}, {Ruhl}, {Runyan}, \& {Torbet}}]{acbar03}
{Goldstein}, J.~H., {et~al.} 2003, \apj, 599, 773

\bibitem[{{Grin} \& {Hirata}(2010)}]{Grin2009}
{Grin}, D., \& {Hirata}, C.~M. 2010, \prd, 81, 083005

\bibitem[{{Hirata}(2008)}]{Hirata2008}
{Hirata}, C.~M. 2008, \prd, 78, 023001

\bibitem[{{Hou} {et~al.}(2011){Hou}, {Keisler}, {Knox}, {Millea}, \&
  {Reichardt}}]{hou11}
{Hou}, Z., {Keisler}, R., {Knox}, L., {Millea}, M., \& {Reichardt}, C. 2011,
  ArXiv e-prints

\bibitem[{{Hou} {et~al.}(2012){Hou}, {Reichardt}, {Story}, {Follin}, {Keisler},
  {Aird}, {Benson}, {Bleem}, {Carlstrom}, {Chang}, {Cho}, {Crawford}, {Crites},
  {de Haan}, {de Putter}, {Dobbs}, {Dodelson}, {Dudley}, {George}, {Halverson},
  {Holder}, {Holzapfel}, {Hoover}, {Hrubes}, {Joy}, {Knox}, {Lee}, {Leitch},
  {Lueker}, {Luong-Van}, {McMahon}, {Mehl}, {Meyer}, {Millea}, {Mohr},
  {Montroy}, {Padin}, {Plagge}, {Pryke}, {Ruhl}, {Sayre}, {Schaffer}, {Shaw},
  {Shirokoff}, {Spieler}, {Staniszewski}, {Stark}, {van Engelen},
  {Vanderlinde}, {Vieira}, {Williamson}, \& {Zahn}}]{hou12}
{Hou}, Z., {et~al.} 2012, ArXiv e-prints

\bibitem[{{Hu} {et~al.}(1995){Hu}, {Scott}, {Sugiyama}, \& {White}}]{Hu1995}
{Hu}, W., {Scott}, D., {Sugiyama}, N., \& {White}, M. 1995, \prd, 52, 5498

\bibitem[{{Hu} \& {White}(1997)}]{hu97}
{Hu}, W., \& {White}, M. 1997, \apj, 479, 568

\bibitem[{{H{\"u}tsi} {et~al.}(2011){H{\"u}tsi}, {Chluba}, {Hektor}, \&
  {Raidal}}]{hue11}
{H{\"u}tsi}, G., {Chluba}, J., {Hektor}, A., \& {Raidal}, M. 2011, \aap, 535,
  A26

\bibitem[{{H{\"u}tsi} {et~al.}(2009){H{\"u}tsi}, {Hektor}, \& {Raidal}}]{hue09}
{H{\"u}tsi}, G., {Hektor}, A., \& {Raidal}, M. 2009, \aap, 505, 999

\bibitem[{{Kaiser}(1983)}]{kai83}
{Kaiser}, N. 1983, \mnras, 202, 1169

\bibitem[{{Kaplinghat} {et~al.}(1999){Kaplinghat}, {Scherrer}, \&
  {Turner}}]{Kaplinghat1999}
{Kaplinghat}, M., {Scherrer}, R.~J., \& {Turner}, M.~S. 1999, \prd, 60, 023516

\bibitem[{{Keisler} {et~al.}(2011){Keisler}, {Reichardt}, {Aird}, {Benson},
  {Bleem}, {Carlstrom}, {Chang}, {Cho}, {Crawford}, {Crites}, {de Haan},
  {Dobbs}, {Dudley}, {George}, {Halverson}, {Holder}, {Holzapfel}, {Hoover},
  {Hou}, {Hrubes}, {Joy}, {Knox}, {Lee}, {Leitch}, {Lueker}, {Luong-Van},
  {McMahon}, {Mehl}, {Meyer}, {Millea}, {Mohr}, {Montroy}, {Natoli}, {Padin},
  {Plagge}, {Pryke}, {Ruhl}, {Schaffer}, {Shaw}, {Shirokoff}, {Spieler},
  {Staniszewski}, {Stark}, {Story}, {van Engelen}, {Vanderlinde}, {Vieira},
  {Williamson}, \& {Zahn}}]{spt}
{Keisler}, R., {et~al.} 2011, \apj, 743, 28

\bibitem[{{Kholupenko} \& {Ivanchik}(2006)}]{Kholu2006}
{Kholupenko}, E.~E., \& {Ivanchik}, A.~V. 2006, Astronomy Letters, 32, 795

\bibitem[{{Larson} {et~al.}(2011){Larson}, {Dunkley}, {Hinshaw}, {Komatsu},
  {Nolta}, {Bennett}, {Gold}, {Halpern}, {Hill}, {Jarosik}, {Kogut}, {Limon},
  {Meyer}, {Odegard}, {Page}, {Smith}, {Spergel}, {Tucker}, {Weiland},
  {Wollack}, \& {Wright}}]{wmap7}
{Larson}, D., {et~al.} 2011, \apjs, 192, 16

\bibitem[{{Lewis} {et~al.}(2000){Lewis}, {Challinor}, \& {Lasenby}}]{CAMB}
{Lewis}, A., {Challinor}, A., \& {Lasenby}, A. 2000, \apj, 538, 473

\bibitem[{{Lewis} {et~al.}(2006){Lewis}, {Weller}, \& {Battye}}]{Lewis2006}
{Lewis}, A., {Weller}, J., \& {Battye}, R. 2006, \mnras, 373, 561

\bibitem[{{Lizarraga} {et~al.}(2012){Lizarraga}, {Sendra}, \&
  {Urrestilla}}]{liz12}
{Lizarraga}, J., {Sendra}, I., \& {Urrestilla}, J. 2012, \prd, 86, 123014

\bibitem[{{Monaghan}(2005)}]{mon05}
{Monaghan}, J.~J. 2005, Reports on Progress in Physics, 68, 1703

\bibitem[{{Mortonson} {et~al.}(2010){Mortonson}, {Huterer}, \& {Hu}}]{mor10}
{Mortonson}, M.~J., {Huterer}, D., \& {Hu}, W. 2010, \prd, 82, 063004

\bibitem[{{Padmanabhan} \& {Finkbeiner}(2005)}]{pad05}
{Padmanabhan}, N., \& {Finkbeiner}, D.~P. 2005, \prd, 72, 023508

\bibitem[{{Peebles}(1968)}]{Peebles68}
{Peebles}, P.~J.~E. 1968, \apj, 153, 1

\bibitem[{{Peebles} {et~al.}(2000){Peebles}, {Seager}, \& {Hu}}]{Peebles2000}
{Peebles}, P.~J.~E., {Seager}, S., \& {Hu}, W. 2000, \apjl, 539, L1

\bibitem[{{Reichardt} {et~al.}(2009){Reichardt}, {Ade}, {Bock}, {Bond},
  {Brevik}, {Contaldi}, {Daub}, {Dempsey}, {Goldstein}, {Holzapfel}, {Kuo},
  {Lange}, {Lueker}, {Newcomb}, {Peterson}, {Ruhl}, {Runyan}, \&
  {Staniszewski}}]{acbar09}
{Reichardt}, C.~L., {et~al.} 2009, \apj, 694, 1200

\bibitem[{{Rubi{\~n}o-Mart{\'{\i}}n} {et~al.}(2010){Rubi{\~n}o-Mart{\'{\i}}n},
  {Chluba}, {Fendt}, \& {Wandelt}}]{rub10}
{Rubi{\~n}o-Mart{\'{\i}}n}, J.~A., {Chluba}, J., {Fendt}, W.~A., \& {Wandelt},
  B.~D. 2010, \mnras, 403, 439

\bibitem[{{Samsing} {et~al.}(2012){Samsing}, {Linder}, \& {Smith}}]{sam12}
{Samsing}, J., {Linder}, E.~V., \& {Smith}, T.~L. 2012, ArXiv:1208.4845

\bibitem[{{Sc{\'o}ccola} {et~al.}(2008){Sc{\'o}ccola}, {Landau}, \&
  {Vucetich}}]{Scoccola2008}
{Sc{\'o}ccola}, C.~G., {Landau}, S.~J., \& {Vucetich}, H. 2008, Physics Letters
  B, 669, 212

\bibitem[{{Seager} {et~al.}(1999){Seager}, {Sasselov}, \& {Scott}}]{sea99}
{Seager}, S., {Sasselov}, D.~D., \& {Scott}, D. 1999, \apjl, 523, L1

\bibitem[{{Seljak} \& {Zaldarriaga}(1996)}]{CMBFAST}
{Seljak}, U., \& {Zaldarriaga}, M. 1996, \apj, 469, 437

\bibitem[{{Sendra} \& {Smith}(2012)}]{sen12}
{Sendra}, I., \& {Smith}, T.~L. 2012, \prd, 85, 123002

\bibitem[{{Shaw} \& {Chluba}(2011)}]{sha11}
{Shaw}, J.~R., \& {Chluba}, J. 2011, \mnras, 415, 1343

\bibitem[{{Sievers} {et~al.}(2003){Sievers}, {Bond}, {Cartwright}, {Contaldi},
  {Mason}, {Myers}, {Padin}, {Pearson}, {Pen}, {Pogosyan}, {Prunet},
  {Readhead}, {Shepherd}, {Udomprasert}, {Bronfman}, {Holzapfel}, \&
  {May}}]{cbi03}
{Sievers}, J.~L., {et~al.} 2003, \apj, 591, 599

\bibitem[{{Sievers} {et~al.}(2007){Sievers}, {Achermann}, {Bond}, {Bronfman},
  {Bustos}, {Contaldi}, {Dickinson}, {Ferreira}, {Jones}, {Lewis}, {Mason},
  {May}, {Myers}, {Oyarce}, {Padin}, {Pearson}, {Pospieszalski}, {Readhead},
  {Reeves}, {Taylor}, \& {Torres}}]{cbi07}
---. 2007, \apj, 660, 976

\bibitem[{{Sievers} {et~al.}(2013){Sievers}, {Hlozek}, {Nolta}, {Acquaviva},
  {Addison}, {Ade}, {Aguirre}, {Amiri}, {Appel}, {Barrientos}, {Battistelli},
  {Battaglia}, {Bond}, {Brown}, {Burger}, {Calabrese}, {Chervenak}, {Crichton},
  {Das}, {Devlin}, {Dicker}, {Bertrand Doriese}, {Dunkley}, {D{\"u}nner},
  {Essinger-Hileman}, {Faber}, {Fisher}, {Fowler}, {Gallardo}, {Gordon},
  {Gralla}, {Hajian}, {Halpern}, {Hasselfield}, {Hern{\'a}ndez-Monteagudo},
  {Hill}, {Hilton}, {Hilton}, {Hincks}, {Holtz}, {Huffenberger}, {Hughes},
  {Hughes}, {Infante}, {Irwin}, {Jacobson}, {Johnstone}, {Baptiste Juin},
  {Kaul}, {Klein}, {Kosowsky}, {Lau}, {Limon}, {Lin}, {Louis}, {Lupton},
  {Marriage}, {Marsden}, {Martocci}, {Mauskopf}, {McLaren}, {Menanteau},
  {Moodley}, {Moseley}, {Netterfield}, {Niemack}, {Page}, {Page}, {Parker},
  {Partridge}, {Plimpton}, {Quintana}, {Reese}, {Reid}, {Rojas}, {Sehgal},
  {Sherwin}, {Schmitt}, {Spergel}, {Staggs}, {Stryzak}, {Swetz}, {Switzer},
  {Thornton}, {Trac}, {Tucker}, {Uehara}, {Visnjic}, {Warne}, {Wilson},
  {Wollack}, {Zhao}, \& {Zuncke}}]{sie13}
---. 2013, ArXiv e-prints

\bibitem[{{Silk}(1968)}]{sil68}
{Silk}, J. 1968, \apj, 151, 459

\bibitem[{{Singh} {et~al.}(2003){Singh}, {Misra}, {Misra}, {Hnizdo},
  {Fedorowicz}, \& {Demchuk}}]{sin03}
{Singh}, H., {Misra}, N., {Misra}, N., {Hnizdo}, V., {Fedorowicz}, A., \&
  {Demchuk}, E. 2003, American Journal of Mathematical and Management Sciences,
  23, 301

\bibitem[{{Smith} {et~al.}(2012){Smith}, {Das}, \& {Zahn}}]{smi12}
{Smith}, T.~L., {Das}, S., \& {Zahn}, O. 2012, \prd, 85, 023001

\bibitem[{{Smith} {et~al.}(2006){Smith}, {Pierpaoli}, \&
  {Kamionkowski}}]{smi06}
{Smith}, T.~L., {Pierpaoli}, E., \& {Kamionkowski}, M. 2006, Physical Review
  Letters, 97, 021301

\bibitem[{{Story} {et~al.}(2012){Story}, {Reichardt}, {Hou}, {Keisler}, {Aird},
  {Benson}, {Bleem}, {Carlstrom}, {Chang}, {Cho}, {Crawford}, {Crites}, {de
  Haan}, {Dobbs}, {Dudley}, {Follin}, {George}, {Halverson}, {Holder},
  {Holzapfel}, {Hoover}, {Hrubes}, {Joy}, {Knox}, {Lee}, {Leitch}, {Lueker},
  {Luong-Van}, {McMahon}, {Mehl}, {Meyer}, {Millea}, {Mohr}, {Montroy},
  {Padin}, {Plagge}, {Pryke}, {Ruhl}, {Sayre}, {Schaffer}, {Shaw}, {Shirokoff},
  {Spieler}, {Staniszewski}, {Stark}, {van Engelen}, {Vanderlinde}, {Vieira},
  {Williamson}, \& {Zahn}}]{spt12a}
{Story}, K.~T., {et~al.} 2012, ArXiv e-prints

\bibitem[{{Sunyaev} \& {Chluba}(2009)}]{Sunyaev2009}
{Sunyaev}, R.~A., \& {Chluba}, J. 2009, Astronomische Nachrichten, 330, 657

\bibitem[{{Switzer} \& {Hirata}(2008)}]{Switzer2007I}
{Switzer}, E.~R., \& {Hirata}, C.~M. 2008, \prd, 77, 083006

\bibitem[{{Tegmark} {et~al.}(1997){Tegmark}, {Taylor}, \& {Heavens}}]{teg97}
{Tegmark}, M., {Taylor}, A.~N., \& {Heavens}, A.~F. 1997, \apj, 480, 22

\bibitem[{{Verde} {et~al.}(2003){Verde}, {Peiris}, {Spergel}, {Nolta},
  {Bennett}, {Halpern}, {Hinshaw}, {Jarosik}, {Kogut}, {Limon}, {Meyer},
  {Page}, {Tucker}, {Wollack}, \& {Wright}}]{ver03}
{Verde}, L., {et~al.} 2003, \apjs, 148, 195

\bibitem[{{Vikhlinin} {et~al.}(2009){Vikhlinin}, {Kravtsov}, {Burenin},
  {Ebeling}, {Forman}, {Hornstrup}, {Jones}, {Murray}, {Nagai}, {Quintana}, \&
  {Voevodkin}}]{vik09}
{Vikhlinin}, A., {et~al.} 2009, \apj, 692, 1060

\bibitem[{{Zeldovich} {et~al.}(1968){Zeldovich}, {Kurt}, \&
  {Syunyaev}}]{Zeldovich68}
{Zeldovich}, Y.~B., {Kurt}, V.~G., \& {Syunyaev}, R.~A. 1968, Zhurnal
  Eksperimental noi i Teoreticheskoi Fiziki, 55, 278

\bibitem[{{Zhang} {et~al.}(2007){Zhang}, {Chen}, {Kamionkowski}, {Si}, \&
  {Zheng}}]{zha07}
{Zhang}, L., {Chen}, X., {Kamionkowski}, M., {Si}, Z., \& {Zheng}, Z. 2007,
  \prd, 76, 061301

\bibitem[{{Zhang} {et~al.}(2006){Zhang}, {Chen}, {Lei}, \& {Si}}]{zha06}
{Zhang}, L., {Chen}, X., {Lei}, Y., \& {Si}, Z. 2006, \prd, 74, 103519

\end{thebibliography}
\bibliographystyle{apj}
\clearpage

\end{document}